\documentclass[%
 reprint,
 amsmath,amssymb,
 aps,
 pra,
 longbibliography,
 nofootinbib
]{revtex4-2}

\usepackage{hyperref}
\usepackage{graphicx}
\usepackage{dcolumn}
\usepackage{bm}
\usepackage{xcolor}
\usepackage{siunitx}
\usepackage{soul}
\usepackage[bottom]{footmisc}% places footnotes at page bottom
\usepackage{url} % Use the package "url.sty" to avoid problems with special characters used in your e-mail or web address

\usepackage{braket}

\begin{document}

\preprint{APS/123-QED}
% Title
\title{Mediated interactions in mixtures of ultracold atoms}
\author{Henry Ando}
\author{Geyue Cai}
\author{Cheng Chin}
\affiliation{The James Franck Institute, Enrico Fermi Institute, and Department of Physics, The University of Chicago, Chicago, IL, USA}
\author{Tilman Enss}
\affiliation{Institute for Theoretical Physics, University of Heidelberg, Heidelberg, Germany}

\begin{abstract}
We describe recent theoretical and experimental developments on mediated interactions in mixtures of bosonic and fermionic atoms. We discuss how particle-hole excitations of a Fermi sea can induce long-range interactions between heavy impurities or atoms in a Bose-Einstein condensate. Conversely, phonon excitations of a Bose-Einstein condensate induce interactions between fermionic atoms. These mediated interactions exhibit different short-range and long-range scaling regimes with distance and, if strong enough, can induce fermion superfluidity. We discuss the prospects for observing new phenomena that could arise from mediated interactions. Experimentally, we outline recent studies of the $^{133}$Cs–$^6$Li Bose-Fermi mixture, a platform well-suited for investigating fermion-mediated interactions. A Cs Bose–Einstein condensate immersed in a degenerate Li Fermi gas is prepared with tunable interspecies interactions. In the weak-coupling regime, precision measurements of condensate properties reveal fermion-mediated attractions between bosons, matching theoretical predictions. In the strong-coupling regime, we observe suppression and revival of sound modes and novel many-body resonances. Altogether, we aim to highlight both instances where experiment and theory agree well, and promising prospects to engineer long-range interactions in atomic quantum gases.
\end{abstract}

\maketitle

\section{Introduction}

Ultracold mixtures of fermionic and bosonic atoms exhibit some of the most interesting many- and few-body effects, including superfluidity \cite{wang2006,ferrier2014,roy2017,kinnunen2018,bighin2025}, quantum phase transitions \cite{sachdev1999,powell2005}, Casimir interactions \cite{nishida2009casimir,enss2020scattering} and the Efimov effect \cite{efimov1970,naidon2017,greene2017}, in a single system. They can be modeled by a Hamiltonian
\begin{align}
    \label{eq:ham}
    H =& \sum_{i=1}^{N_\text{F}} \frac{\vec p_{\text{F}i}^2}{2m_\text{F}}
    + \sum_{i=1}^{N_\text{B}} \frac{\vec p_{\text{B}i}^2}{2m_\text{B}} \nonumber\\
    &+ \sum_{i<j} V_\text{BF}(\vec r_{\text{B}i}-\vec r_{\text{F}j}) + \sum_{i<j} V_\text{BB}(\vec r_{\text{B}i}-\vec r_{\text{B}j}),
\end{align}
where $\vec p_{\text{F/B}i}$ are the momenta of the $i-$th fermionic and bosonic atoms, $\vec r_{\text{F/B}i}$ denote their positions, $m_\text{F/B}$ are their masses, and $N_\text{F/B}$ are the total numbers of each species. The interactions between bosons and fermions are typically of much shorter range than the particle spacing and well approximated by a 3D Fermi pseudopotential, $V_\text{BF}(\vec r) \psi(\vec r) = g_\text{BF} \delta^{(3)}(\vec r) \frac{d}{dr}[r\psi(r)]$ \cite{bloch2008}.  The 3D boson-fermion coupling strength $g_\text{BF} = 2\pi\hbar^2a_\text{BF}/\mu_\text{BF}$ is given in terms of the experimentally tunable \cite{chin2010feshbach} boson-fermion scattering length $a_\text{BF}$ and reduced mass $\mu_\text{BF} =m_\text{F}m_\text{B}/(m_\text{F}+m_\text{B})$.  The Bose-Bose interaction $V_\text{BB}$ is described analogously, while the Pauli principle prevents a zero-range interaction between identical fermions (generalizations with several fermion species will also have $V_\text{FF}$ \cite{wang2006}).

At weak coupling and low temperature the bosons form a Bose-Einstein condensate (BEC) mixed with a Fermi gas. At strong Bose-Fermi attraction, bosons and fermions can bind into molecules of fermionic statistics, which no longer condense.  The point where the BEC is fully converted into molecules marks a quantum phase transition \cite{powell2005, duda2023}.  The stability of the mixture depends delicately on the interactions between bosons and fermions mediated by the surrounding medium.  A repulsive mediated interaction can lead to phase separation \cite{molmer1998,viverit2000,lous2018}, while an attractive mediated interaction might lead to collapse or droplet formation \cite{molmer1998,marchetti2008,rakshit2019}. These mediated interactions arise from the complex interplay between statistics, few-body bound states and many-body medium excitations. A fermionic medium induces an interaction that oscillates between repulsive and attractive as a function of the interatomic distance \cite{ruderman1954,kasuya1956,yosida1957}: this fermion mediated interaction was first observed for localized moments in metals and is now realized in fermionic quantum gases \cite{nishida2009casimir,desalvo2019,edri2020,chen2022}. Generally speaking, lighter medium particles induce a stronger mediated interaction between the impurities \cite{enss2020scattering}.

In our exposition we aim not for completeness (see also the recent reviews \cite{paredes2024, grusdt2025, massignan2025}) but instead focus on explaining the fundamental mechanisms that give rise to mediated interactions and illustrate them with recent experimental observations.  Therefore, we start our discussion with the impurity limits: few bosonic impurities in a large Fermi sea (Sec.~\ref{sec:fermi}), or few impurities in a BEC (Sec.~\ref{sec:bec}); these limits are tractable but nontrivial (in some cases even the sign of mediated interactions depends on the way it is experimentally probed \cite{levinsen2025medium}).  The observation of mediated interactions and the interplay between few- and many-body effects is illustrated with experimental measurements in Cs-Li mixtures (Sec.~\ref{sec:lics}).  We conclude with an outlook in Sec.~\ref{sec:out}.

\section{Interactions mediated by fermions}
\label{sec:fermi}

In order to understand how particle-hole excitations of a fermionic medium can induce interactions between another species of atoms, consider first the simplest case of just two ``impurity'' atoms inserted into a Fermi sea.  How does the energy of this configuration change with the distance between the impurities?  For two heavy impurities this is an example of the Casimir effect: the impurities provide a boundary condition on the fermion wave functions, which constrains the quantum and thermal fluctuations of the medium and changes the zero-point energy.  From the perspective of the impurities, the dependence of the energy on their separation $R$ acts as a mediated interaction potential $V(R)$.

This Casimir interaction is easily understood for the case of two infinitely heavy (fixed) impurities at separation $R$ that are interacting with a fermionic medium.  The short-range interaction imposes the Bethe-Peierls boundary condition on the fermionic wave functions at position $\vec r$,
\begin{align}
    \label{eq:bethe}
    \psi(\vec r\to \vec R_i) \propto \frac1{|\vec r-\vec R_i|}-\frac1{a_\text{BF}}+\mathcal O(|\vec r-\vec R_i|),
\end{align}
near each impurity position $\vec R_i$.  While a fermion can form a bound state with a single impurity only for strong attraction $1/a_\text{BF}>0$ beyond the Feshbach resonance, remarkably it can form a bound state $\kappa_+$ with \emph{two} impurities already at weaker attraction $1/a_\text{BF} > -1/R$; a second bound state $\kappa_-$ appears at stronger attraction for $1/a_\text{BF} > 1/R$, with bound state energies $-\kappa_\pm^2/2m_\text{F}<0$ \cite{nishida2009casimir, enss2020scattering}.  The continuum of fermionic scattering states is characterized by phase shifts $\tan\delta_\pm(k,R) = -\frac{kR\pm\sin(kR)}{R/a\pm\cos(kR)}$ \cite{nishida2009casimir, enss2020scattering}.  The energy change of the occupied fermion states in the presence of the impurities follows by Fumi's theorem \cite{fumi1955},
\begin{align}
    \label{eq:fumi}
    \Delta E(R) =& -\frac{\kappa_+^2(R) + \kappa_-^2(R)}{2m_\text{F}} \nonumber\\
    &- \int_0^{k_\text{F}} dk\, k\, \frac{\delta_+(k,R) + \delta_-(k,R)}{\pi m_\text{F}} \,,
\end{align}
where $\kappa_\pm(R) = 1/a+W(\pm e^{-R/a})/R$ with the Lambert $W$ function, and $k_\text{F}$ denotes the Fermi wave vector.  The mediated two-body interaction is then obtained as the energy shift at separation $R$,
\begin{align}
    \label{eq:pot}
    V(R) = \Delta E(R) - \Delta E(R\to\infty),
\end{align}
with the single-particle energies of two infinitely separated impurities subtracted. 
The resulting interaction $V(R)$ shows strongly attractive Efimov scaling for short distance and oscillatory behavior for distances larger than the fermion spacing; we now discuss both in turn.

\begin{figure*}
    \centering
    \includegraphics[width=0.75\linewidth]{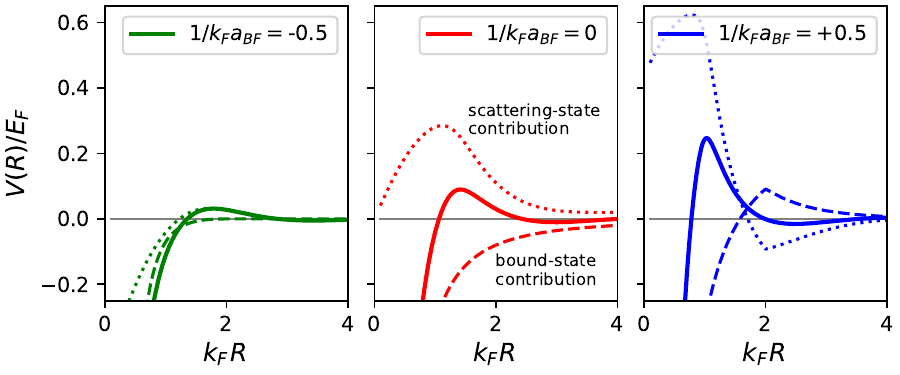}
    \caption{Mediated interaction $V(R)$ between bosons in a Fermi gas for different Bose-Fermi coupling $a_\text{BF}$; the interaction is normalized by the Fermi energy $E_\text{F}$.  The contribution from scattering states (RKKY; dotted) combines with the bound-state contribution (dashed) to the full interaction (solid lines). For repulsive $a_\text{BF}>0$ the bound-state contribution reaches positive values because it is measured from the single-particle energy, Eq.~\eqref{eq:pot}.}
    \label{fig:medint}
    % ALTTEXT: Three-panel chart showing the mediated interaction potential energy V(R)/E_F versus k_F R for different values of the interaction parameter 1/k_F a_{BF}. - **Left panel**: At negative scattering length 1/k_F a_{BF} = -0.5, the interaction peaks around k_F R = 2. - **Middle panel**: At resonant scattering 1/k_F a_{BF} = 0, there is a contribution from scattering states that is positive, shown as the dotted curve. The contribution from bound states is negative, shown as the dashed curve. The total interaction, indicated by the solid line, has a peak around k_F R = 1.5. - **Right panel**: For repulsive scattering 1/k_F a_{BF} = +0.5, there are similar contributions as in the middle panel, with the total interaction peaking around k_F R = 1.
\end{figure*}

\subsection{Two impurities and a single fermion: Efimov effect}
\label{sec:fermi:single}

Even in the case of an extremely dilute fermionic medium with only a single fermion, the lowest bound state $\kappa_+$ will be occupied for $1/a_\text{BF}>-1/R$ and gives rise to a strongly attractive potential at short distances $R$ \cite{nishida2009casimir, enss2020scattering},
\begin{align}
    \label{eq:efimov}
    V^\text{Efimov}(R) = -\frac{\kappa_+^2(R)}{2m_\text{F}} \;
    \xrightarrow{R\to0} \; -\frac{c^2}{2m_\text{F}R^2}\,.
\end{align}
At resonance $1/a_\text{BF}=0$ one finds $\kappa_+=c/R$ at all distances with $c\approx 0.567$ (the bound-state contribution to the potential is shown as the  dashed line in Fig.~\ref{fig:medint}, center panel).  For attractive $a_\text{BF}<0$ the bound-state potential is limited to $R<|a_\text{BF}|$ and vanishes for larger distance (see left panel).  For repulsive interaction $a_\text{BF}>0$ the second bound state $\kappa_-$ starts to appear for distances $R>a_\text{BF}$, marked by the kink in the bound-state potential (right panel).  The $-1/R^2$ potential exhibits classical scale invariance, however quantum fluctuations break this down to a discrete scale invariance, which gives rise to an infinite tower of three-body (impurity-impurity-fermion) bound states.  These Efimov bound states are known from nuclear physics and exhibit characteristic scaling relations \cite{efimov1970, naidon2017}.  A short-distance cutoff, comparable to the van der Waals length, defines a unique, lowest-energy Efimov state of smallest size, followed by an infinite tower of subsequently larger states.  As the fermion density is increased, however, the attractive potential is effectively screened for distances $R$ larger than the fermion spacing, and it therefore becomes harder to form Efimov states of size larger than the fermion spacing, as we will now see.

\subsection{Two impurities in a Fermi sea: RKKY interaction}
\label{sec:fermi:rkky}

For two impurities in a dense Fermi gas, the contribution of scattering states in Eq.~\eqref{eq:fumi} yields the Ruderman-Kittel-Kasuya-Yosida (RKKY) potential, which was first obtained in condensed-matter physics for localized magnetic moments in a metal \cite{ruderman1954, kasuya1956, yosida1957}.  At weak coupling  it reads to quadratic order $\mathcal O(a_\text{BF}^2)$
\begin{align}
    \label{eq:RKKY}
    V^\text{RKKY}(R) = \frac{a_\text{BF}^2}{2m_\text{F}}
    \frac{2k_\text{F} R \cos(2k_\text{F} R) - \sin(2k_\text{F} R)}
    {2\pi R^4}.
\end{align}
The RKKY interaction scales as $-1/R$ for short distances $R\gtrsim |a_\text{BF}|$ and as $\cos(2k_\text{F}R)/R^3$ for $R\to\infty$.  The oscillation arises from the sharp Fermi edge in the momentum distribution at zero temperature.  The interaction can be viewed as one impurity creating Friedel oscillations \cite{friedel1958} in the density and the other one probing them.  At stronger coupling the mediated interaction is obtained numerically from Eq.~\eqref{eq:fumi} and shown in Fig.~\ref{fig:medint}: it has a qualitatively similar oscillatory form and the height of the repulsive barrier, which arises from medium states repelled by the two impurities with $\delta_\pm(k)<0$, grows toward the repulsive side of the resonance.  

In distinction to condensed matter settings, ultracold mixtures with short-range attraction exhibit also the bound-state contribution from Sec.~\ref{sec:fermi:single}, which leads to stronger short-range mediated attraction than the pure RKKY potential.  On the other hand, the repulsive barrier of the RKKY interaction inhibits the formation of Efimov bound states that are larger than the fermion spacing, truncating the infinite tower of bound states to the first few ones \cite{bulgac2001, sun2019efimov, tran2021}.  It is an interesting question whether bound states in the attractive regime $a_\text{BF}<0$ arise from the Efimov part of the interaction, with a single fermion bound to two impurities, or from the RKKY part with delocalized fermionic scattering states; see Fig.~\ref{fig:mediated_resonance}.

\begin{figure}
    \centering
    \includegraphics[width=.9\linewidth]{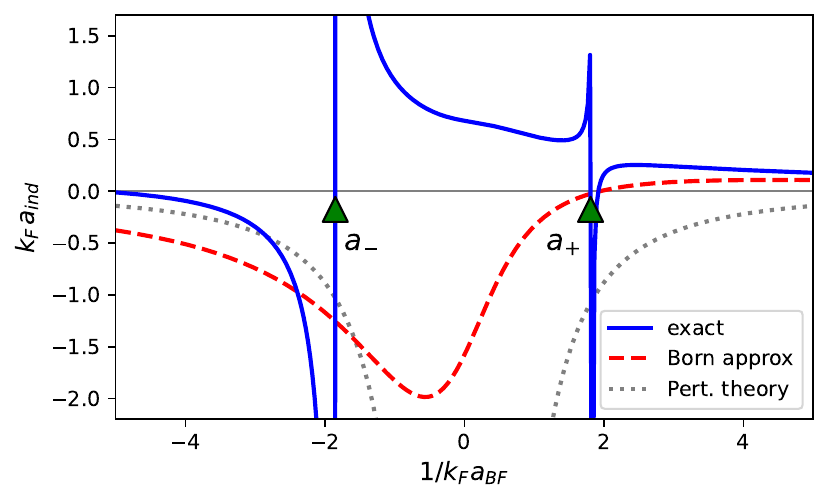}
    \caption{Induced scattering length $a_\text{ind}(a_\text{BF})$ between two heavy impurities in a Fermi sea.  The full solution of the Schr\"odinger equation Eq.~\eqref{eq:varphase} [blue solid] supports resonant scattering between the impurities [arrows]. It is compared to the Born approximation Eq.~\eqref{eq:born} [red dashed] and perturbation theory Eq.~\eqref{eq:aind} [gray dotted], which show no induced resonance.}
    \label{fig:aindth}
    % ALTTEXT: Graph showing three curves for the induced scattering length between two impurities as a function of the inverse Bose-Fermi scattering length. The blue solid line represents the exact solution, which features two resonances, as indicated by the arrows, where the induced scattering changes sign. The red dashed line represents the Born approximation, which is mostly attractive. The gray dotted line represents perturbation theory, which is always attractive and diverges at unitarity.
\end{figure}

\subsection{Induced impurity-impurity scattering}

When the impurities are not infinitely heavy but can move slowly (Born-Oppenheimer approximation), one can consider the scattering between impurities subject to the mediated interaction $V(R)$.  Within weak-coupling perturbation theory the induced scattering length $a_\text{ind}$ between two impurities can be computed to second order in $a_\text{BF}$.  The result is always attractive both for positive and negative $a_\text{BF}$ \cite{santamore2008}, see the gray dotted line in Fig.~\ref{fig:aindth}:
\begin{align}\label{eq:aind}
    a_\text{ind,PT}
    & = - \frac{k_\text{F}}{2\pi} \frac{(m_\text{B}+m_\text{F})^2}{m_\text{B}m_\text{F}}a_\text{BF}^2 \, .
\end{align}
This perturbative expression diverges at strong coupling.  A finite result is obtained within first-order Born approximation \cite{enss2020scattering}, which yields
\begin{align}\label{eq:born}
    a_\text{ind,Born} = m_\text{B}\int_0^\infty dR\, R^2\,V(R)\,,
\end{align}
see the red dashed line in Fig.~\ref{fig:aindth}.  For weak coupling the induced scattering length arising from the RKKY potential Eq.~\eqref{eq:RKKY} agrees with the perturbative result Eq.~\eqref{eq:aind} for large mass ratio $m_\text{B}\gg m_\text{F}$, as realized, e.g., in Cs-Li mixtures.  Interestingly, at weak attraction $a_\text{BF}<0$ the bound-state potential Eq.~\eqref{eq:efimov} gives a nonperturbative contribution to $a_\text{ind,Born}$ already at \emph{first} order in $a_\text{BF}$, which dominates over the RKKY interaction in this regime \cite{enss2020scattering}:
\begin{align}
    \label{eq:aindbound}
    a_\text{ind,Born}^\text{bound}
    & = \frac{m_\text{B}}{m_\text{F}}\,\frac{\gamma\, a_\text{BF}}2\,,
\end{align}
with $\gamma\approx0.1$. At stronger coupling the Born approximation is insufficient and one has to solve the Schr\"odinger equation for the two impurities, $[-\hbar^2\nabla_{\vec R}^2/m_\text{B} + \Delta E(R)]\psi(\vec R)=E\psi(\vec R)$, to extract the scattering phase shift $\delta_\ell^\text{ind}(k)$ and from its effective range expansion the induced scattering length $a_\text{ind}$ between two impurities \cite{enss2020scattering}.  Equivalently, $a_\text{ind}$ is obtained directly using the variable phase equation \cite{calogero1967},
\begin{align}\label{eq:varphase}
    \partial_R a_\text{ind}(R) = m_\text{B} V(R) [R-a_\text{ind}(R)]^2,
\end{align}
with initial condition $a_\text{ind}(R_0)=R_0$ for short-distance cutoff $R_0$ and integrating up to the final $a_\text{ind} = a_\text{ind}(R\to\infty)$.  The resulting induced scattering length is shown in Fig.~\ref{fig:aindth}.  Near resonance, the exact result (blue line) differs dramatically from the Born approximation (red) and second-order perturbation theory (gray): at $a_-$ an Efimov bound state at the scattering threshold (indicated by the arrow) leads to a resonance in $a_\text{ind}$.  Beyond this resonance, the induced scattering changes sign and is \emph{repulsive}, which might help to stabilize Bose-Fermi mixtures at strong coupling.  The mediated interaction can also give rise to entanglement between impurities \cite{mistakidis2019repulsive}.

An interesting question is how these results generalize to the case of a finite density of mobile bosons, which is experimentally investigated below in Sec.~\ref{sec:lics:strong}.  Especially at strong coupling this poses a challenging theoretical problem.  A diagrammatic calculation of the induced scattering between condensed bosons in a Bose-Fermi mixture \cite{shen2024strongly} yields results remarkably close to the Born approximation (red dashed lines in Fig.~\ref{fig:aindth}), even when the bosons are at finite density and equal mass with the fermions.  It will be interesting to see whether the full resummation of the bosonic interaction beyond Born approximation can also lead to an Efimov resonance and a repulsive scattering length at finite boson density.

\subsection{Two fermion species and fermionic superfluids}
\label{subsec:two-Fermi}

If the medium consists of two fermionic components, such as different hyperfine states, then attractive $s$-wave scattering between the fermions can induce superfluidity within the medium.  At low temperature the dominant excitations of the superfluid are gapless phonons \cite{bloch2008}.  These resemble the phonon excitations in a BEC, however the dispersion relation $E_k=c_s|\vec k|+\dotsm$ beyond the linear term has a different form: it is curved downward for a superfluid of weakly attractive fermions, but curved upward for tightly bound fermions or bosons \cite{kurkjian2017}.  In both cases, these low-momentum phonon excitations mediate an interaction between impurities, as we will discuss below in Sec.~\ref{sec:bec:eft}.  

In fermionic superfluids at higher temperature approaching $T_c$, the energy gap $\Delta(T)$ of pair-breaking excitations is reduced.  Hence, these become easier to excite and are expected to also mediate an interaction.  In addition, already a single impurity in a two-component medium can form new Efimov states \cite{pierce2019} that will affect also the induced interaction between two impurities.

\section{Interactions mediated by a BEC}
\label{sec:bec}

An interacting BEC is characterized by the chemical potential $\mu$ as the typical energy scale, and the healing length $\xi=\hbar/\sqrt{8\pi n a_\text{BB}}$ as the typical length scale, where $n$ denotes the uniform density and $a_\text{BB}$ the boson-boson scattering length in the medium.  Gapless phase fluctuations, or phonons, dominate at large distances $R>\xi$ or low energies $E<\mu$, while density fluctuations have an energy gap of order $\mu$ and dominate at short distances $R\lesssim \xi$ \cite{pitaevskii2003}.

\subsection{Short range: Efimov scaling}
\label{sec:bec:efimov}

At short impurity distances $R\lesssim\xi$ the interplay between phonon and density fluctuations is well captured by the Gross-Pitaevskii equation (GPE) for the medium wave function $\psi(\vec r)$,
\begin{align}
    \label{eq:gpe}
    \left[ -\frac{\hbar^2\nabla^2}{2m_\text{B}} + V_\text{ext}(\vec r) + g_\text{BB}|\psi(\vec r)|^2\right] \psi(\vec r) = \mu \psi(\vec r).
\end{align}
Static impurities are easily included in the external potential $V_\text{ext}(\vec r)$, while a mobile impurity can be treated by the Lee-Low-Pines transformation \cite{lee1953}: in the frame co-moving with the impurity, the impurity appears as a static potential, but in addition it contributes to the kinetic term of the medium particles, and the recoil of the impurity induces an interaction between medium particles of order $1/m_\text{I}$, where $m_\text{I}$ denotes the impurity mass.

In the low-density limit of a single medium boson, the nonlinear Schr\"odinger equation \eqref{eq:gpe} reduces to the linear equation discussed above in Sec.~\ref{sec:fermi:single}.  This leads to the appearance of an Efimov bound state and a mediated short-range potential that scales as
\begin{align}
    \label{eq:becefimov}
    V(R) \propto -\frac1{R^2}
\end{align}
for $R \lesssim \text{min}(|a_\text{BI}|,\xi)$ \cite{drescher2023}.

\subsection{Medium range: Yukawa scaling and Gross-Pitaevskii theory}
\label{sec:bec:gpe}

For a finite-density bosonic bath the nonlinearity in the GPE \eqref{eq:gpe} becomes important to capture collective density and phase fluctuations.  A static impurity, included via the potential $V_\text{ext}(\vec r)$, deforms the surrounding bosonic density profile; a second impurity experiences this modified density as a mediated interaction of Yukawa form \cite{pethick2008},
\begin{align}
    \label{eq:yukawa}
    V(R) \propto -\frac{e^{-\sqrt2R/\xi}}R.
\end{align}
The full numerical solution of the GPE with short-range boson repulsion \cite{drescher2020, drescher2023} refines and quantifies the crossover between the Efimov and Yukawa scaling regimes \cite{naidon2018}.  For strong mediated attraction, bound states of two impurities, or bipolarons, can form \cite{casteels2013, camacho2018bipolarons}: they could be detected as an additional loss feature in the impurity spectrum, and their binding energy can be computed from a Schr\"odinger equation for two impurities in the nonlocal potential induced by the BEC45.  Note that in the limit of an ideal BEC, $\xi\to\infty$, density fluctuations become gapless and the mediated interaction assumes a Newtonian form $V(R) \propto -1/R$ throughout \cite{drescher2023}.

\subsection{Long range: Van-der-Waals scaling and Effective Field Theory}
\label{sec:bec:eft}

While the Yukawa interaction Eq.~\eqref{eq:yukawa} results from a tree-level process where a single phonon is exchanged, the fluctuating medium also induces higher-order loop corrections with a two-phonon exchange \cite{pavlov2019, fujii2022}.  For gapless phonons, this process mediates a long-range power-law interaction of van der Waals (vdW) type \cite{fujii2022},
\begin{align}
\label{eq:vdw}
V_\text{vdW}(R) 
& \propto -\frac1{R^7} \quad \text{($T=0$)}, \\
V_\text{vdW}(R)
& \propto -\frac T{R^6} \quad \text{($T>0$, $R\gtrsim\hbar c_s/k_BT$).}
\end{align}
These power laws resemble the standard photon induced van der Waals potential in the relativistic and non-relativistic cases, respectively.  In a finite-temperature BEC, the crossover between both scaling forms occurs at distances around the thermal length $\hbar c_s/k_BT$, where $c_s$ denotes the speed of sound \cite{fujii2022}.

While the coefficient of this loop correction is suppressed by a power of the medium gas parameter, $(na_\text{BB}^3)^{1/2}$, its power-law decay will always dominate over the exponential decay of the Yukawa potential Eq.~\eqref{eq:yukawa} at sufficiently long distances.  The explicit effective field theory (EFT) calculation \cite{fujii2022} applies equally to superfluids of bosons or of fermion pairs, which both have low-energy phonon excitations.  When applied to a superfluid, strongly interacting Fermi gas, where the healing length can become as short as the particle spacing and the gas parameter large, the crossover between Yukawa and van der Waals interaction is predicted to occur already at distances of a few times the healing length \cite{fujii2022}, or a few $\mu$m, bringing both potentials within reach of direct experimental verification.

\begin{figure}
    \centering
    \includegraphics[width=\linewidth]{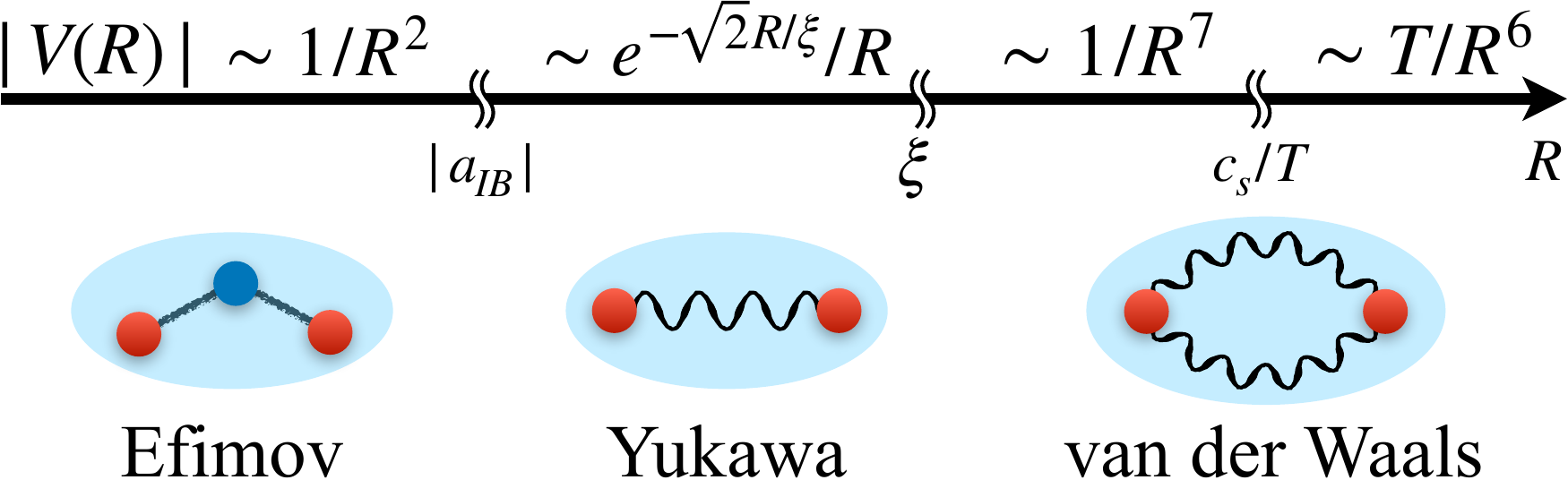}
    \caption{Induced interaction between two heavy impurities in a BEC (red dots): crossover from Efimov scaling Eq.~\eqref{eq:becefimov} at short impurity distances $R$ from scattering off a single boson (blue dot) to Yukawa scaling Eq.~\eqref{eq:yukawa} at intermediate distances from exchanging a single phonon (wavy line) and to van der Waals scaling Eq.~\eqref{eq:vdw} at large distances from exchange of two phonons.}
    \label{fig:aind}
    % ALTTEXT: Mathematical diagram illustrating mediated interactions in a BEC as a function of distance R . For short distances below |a_{IB}| the interaction shows Efimov scaling as 1/R^2; for intermediate distances extending roughly to the healing length it shows Yukawa scaling as e^{-sqrt{2R/xi}/R}. For large distances it shows van-der-Waals scaling as 1/R^7, and beyond the thermal length c_s/T as  T/R^6.
\end{figure}

\subsection{Interaction between fixed vs mobile impurities}

Recent experiments \cite{baroni2024mediated} have found mediated interactions between mobile impurities that differ even in sign from the predictions for heavy impurities.  Why can the impurity mass have such a large effect?  Infinitely heavy impurities remain always localized at a given position; they induce a static deformation of the surrounding medium, whether in the form of Friedel oscillations in a Fermi gas or a monotonically decreasing deformation in a BEC.  The mediated interaction has a clear dependence on the distance between impurities.  Mobile impurities, instead, are often considered not in a localized wave packet but instead in a zero-momentum plane wave state that extends over the whole sample; here the induced interaction is averaged over all distances $R$ between the impurities.  Mobile impurities form quasiparticles whose interaction is described by Landau's theory \cite{camacho2018landau}.

For mobile impurities the quantum statistics of the impurities under exchange has important effects \cite{paredes2024}:  For identical impurities there is an exchange contribution to the mediated interaction, which has opposite sign for bosons and fermions.  This explains the observation in a recent experiment: fermionic $^{40}$K impurities at weak coupling with an ultracold fermionic $^6$Li medium have a repulsive induced interaction, while for bosonic $^{41}$K impurities it is attractive, in agreement with Fermi-liquid theory \cite{baroni2024mediated}.  Note that the contribution of the exchange term also depends on whether the impurities are in identical momentum states, and whether the density or chemical potential is held fixed \cite{levinsen2025medium}.

\subsection{Dense mixtures}

A dense Bose-Fermi mixture, with nonzero density of both bosons and fermions, exhibits a richer phase diagram than the fermionic BCS-BEC crossover \cite{bloch2008}.  At strong attraction, condensed bosons can be bound with fermions into fermionic molecules.  When the boson density is lower than the fermion density, this mechanism can fully deplete the BEC, marking a quantum phase transition at zero temperature \cite{powell2005, manabe2021}.

Mediated interactions in dense Bose-Fermi mixtures have been studied with a view to fermion superfluidity, even in absence of direct fermion attraction: if bosons mediate an attraction between fermions, there is a tendency to form fermion pairs that can lead to superfluidity.  This effect is enhanced for two fermion species ($s$-wave pairing) and when the medium excitations are soft modes that cost little energy.  Calculations within BCS \cite{heiselberg2000}, Eliashberg \cite{wang2006} and Wegner flow theory \cite{enss2009, bighin2025} show that induced superfluidity should be observable up to a few percent of the Fermi temperature. 

\section{Experiments on $^{133}$Cs-$^6$Li Bose-Fermi mixture}\label{sec:lics}

Ultracold mixtures of bosonic and fermionic atoms provide a versatile platform to study novel quantum phenomena that emerge from the interplay between particles obeying different quantum statistics. The mixture of bosonic $^{133}$Cs and fermionic $^{6}$Li is particularly well-suited for studying fermion-mediated interactions between bosons. This is due to the large Cs to Li mass ratio, which offers a clear separation in the bosonic and fermionic energy scales, and also their convenient tunable interspecies interactions.

In this section, we first review the basic interaction properties of the Li-Cs mixture, focusing on the range where the interspecies interactions can be tuned to investigate mediated interactions (Sec.~\ref{sec:lics:coll}). We then give a brief overview of the key experimental steps to prepare and detect the Li-Cs mixture in the normal and quantum gas regimes (Sec.~\ref{sec:lics:prep}). We outline observations of mediated interactions in both the weak-coupling (Sec.~\ref{sec:lics:weak}) and strong-coupling (Sec.~\ref{sec:lics:strong}) regimes, highlighting connections to the theory discussed in Sec.~\ref{sec:fermi} and the prospects to explore new physical phenomena associated with mediated interactions. 

\subsection{Collisional Properties of the $^{133}$Cs-$^{6}$Li Mixture}
\label{sec:lics:coll}

The interspecies interaction between $^{133}$Cs and $^{6}$Li can be conveniently tuned near magnetic Feshbach resonances \cite{repp2013,tung2013}. The resonances provide precise control over the boson-fermion scattering length $a_{\mathrm{BF}}$, enabling exploration of few- and many-body physics from the weakly interacting regime to the strongly correlated unitary limit.

\begin{figure}[htbp]
    \centering
    \includegraphics[width=7cm]{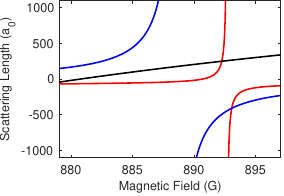}
    \caption{Calculated scattering lengths as a function of magnetic field in the region of interest. The red line shows the narrow Feshbach resonance in the Li$_\text{a}$–Cs spin channel near 893~G, and the blue line indicates a broad resonance in the Li$_\text{b}$–Cs channel near 889~G.  The Cs–Cs scattering length is plotted in black. \textit{Source: Li-Cs scattering lengths extracted from Ref.~\cite{johansen2017} and Cs-Cs scattering lengths extracted from Ref.~\cite{berninger2013}}.}
    \label{fig:feshbach}
\end{figure}

Figure~\ref{fig:feshbach} highlights two particularly useful Li-Cs Feshbach resonances which occur in a range of magnetic fields where a Cs BEC can be prepared. Over this range the Cs-Cs scattering length is small and positive \cite{berninger2013}, which enables the stability of a bare Cs BEC. The narrow Feshbach resonance in the Li$_\text{a}$–Cs spin channel and the broader resonance in the Li$_\text{b}$–Cs channel offer two distinct tuning pathways \cite{repp2013,tung2013}, where Li$_\text{a}$ and Li$_\text{b}$ refer to the lowest and second lowest hyperfine ground state of Li.

\begin{figure*}[htbp]
    \centering
    \includegraphics[width=0.8\textwidth]{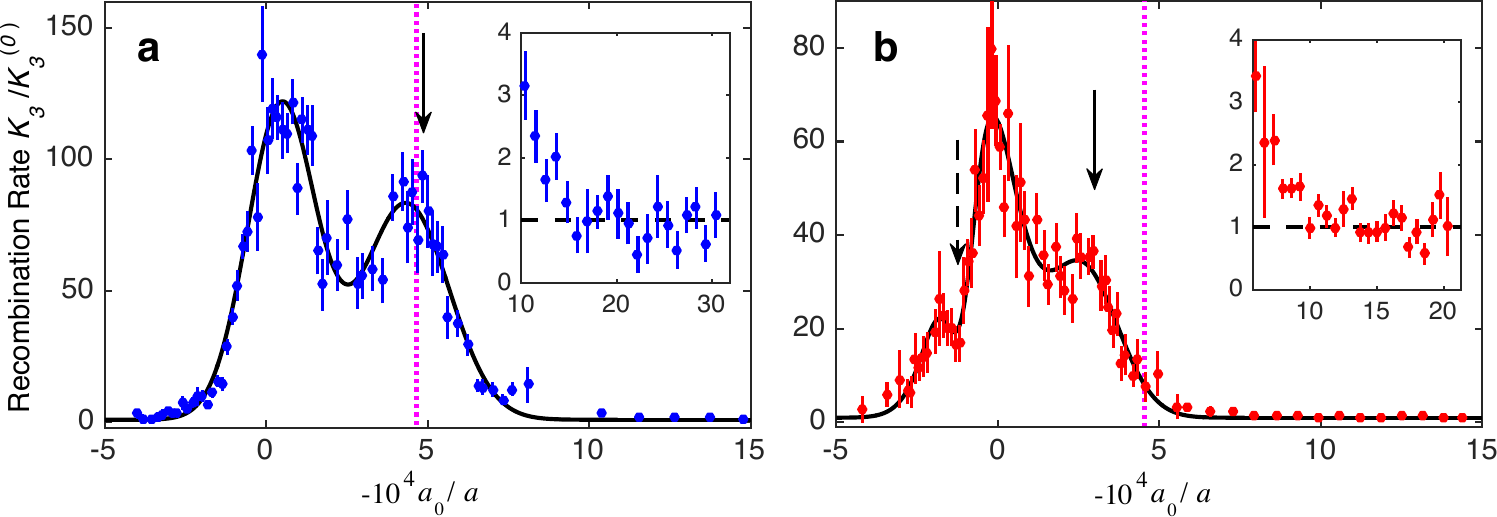}
    \caption{Three-body recombination rate $K_3$ plotted as a function of inverse boson-fermion scattering length $1/a$, demonstrating Efimov resonances near two Li-Cs Feshbach resonances. (\textbf{a}) shows the behavior near the broad Li$_\text{b}$-Cs resonance at 889~G, while (\textbf{b}) focuses on the narrow Li$_\text{a}$-Cs resonance near 893~G. $K_3$ is normalized to the off-resonant value $K_3^{(0)}$. Vertical arrows indicate observed Efimov resonance positions. The dashed arrow in (b) points to a suppression feature in $K_3$ at positive scattering length. Magenta dotted lines show predictions from universal Efimov theory; black curves are guides to the eye consisting of the sum of Gaussians. Insets show data far from the resonance. Error bars represent 1-$\sigma$ statistical uncertainty. \textit{Source: adapted from Ref.~\cite{johansen2017}}.}
    \label{fig:efimov}
\end{figure*}

In addition to tunable two-body interactions, the Cs–Li mixture supports Efimov three-body bound states \cite{tung2014,ulmanis2015,johansen2017}. In such states, one fermion (Li) and two identical bosons (Cs) can form universal trimer states in the presence of strong Li-Cs interactions (see also Sect.~\ref{sec:fermi:single}). Experimentally, Efimov states manifest as resonant enhancement in the three-body recombination rate at specific values of $a_{\mathrm{BF}}$. Fig.~\ref{fig:efimov} shows the observation of Efimov resonances in the thermal mixtures of Li$_\text{a}$-Cs and Li$_\text{b}$-Cs. The Efimov resonances display a geometric scaling of the resonant atomic scattering lengths.

\begin{figure*}[htbp]
    \centering
    \includegraphics[clip, trim=15cm 0cm 0cm 26cm, width=0.8\textwidth]{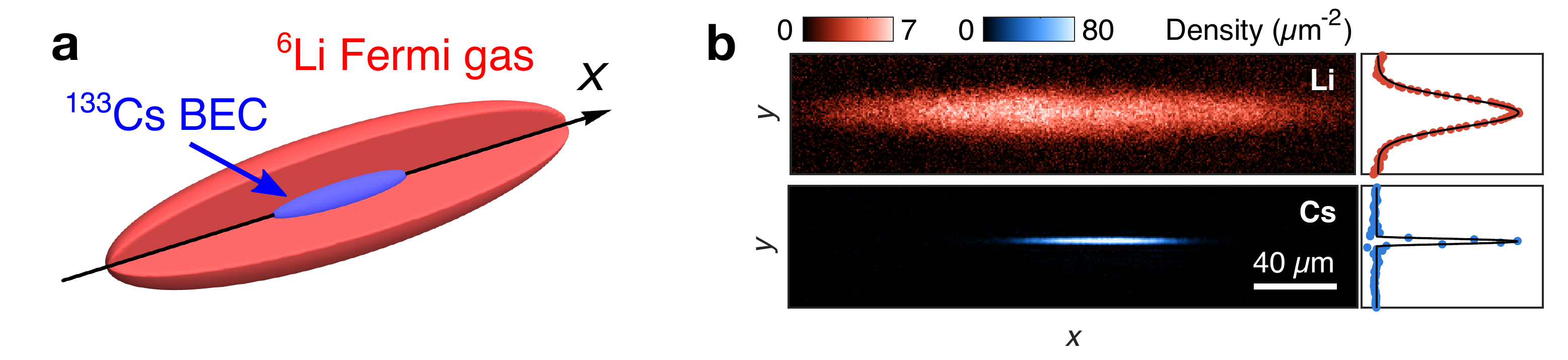}
    \caption{A dual-degenerate Cs-Li Bose-Fermi mixture. (\textbf{a}) Three-dimensional cartoon of the degenerate Cs–Li mixture. The central blue region is a nearly pure $^{133}$Cs BEC, surrounded by a dilute degenerate Fermi gas of $^6$Li. (\textbf{b}) Characteristic \textit{in situ} absorption images of Li (red) and Cs (blue). Integrated density profiles provided at right. \textit{Source: (a) is adapted from Ref.~\cite{patel2023}. (b) is adapted from Ref.~\cite{cai2025}.}}
    \label{fig:csli_image}
\end{figure*}

Altogether, the Feshbach and Efimov resonance structures in the Cs-Li are a key and immutable property of their two- and three-body interactions. They define the accessible parameter space for tuning the microscopic interspecies interactions, creating stable mixtures, Cs-Li molecular energy structure, and exploring the properties of the mixture in the quantum degenerate regime.

\subsection{Preparation of $^{133}$Cs–$^6$Li Quantum Mixture}
\label{sec:lics:prep}

The preparation of a dual-degenerate Bose-Fermi mixture of $^{133}$Cs and $^6$Li involves a multi-stage cooling and trapping sequence tailored to the distinct properties of each species \cite{desalvo2017}. In this section, we seek to highlight key challenges and solutions specific to the mixture, and will gloss over general techniques which are now industry-standard.

Each atomic species is initially cooled by a magneto-optical trap (MOT) and transferred into independent far-off-resonant dipole traps for evaporative cooling. We have observed that loading a MOT of either species near a trapped sample of the other species will lead to heavy loss in the trapped cloud, likely due to light-assisted collisions. Thus, we use a translatable dipole trap to move the trapped Li gas several centimeters away from the Cs MOT before turning it on. 

Another challenge to reaching dual degeneracy is maintaining good overlap between the species during the last stages of evaporative cooling. Initially, Li can be evaporated by itself by using a strongly interacting mixture of Li$_\text{a}$ and Li$_\text{b}$. However, once the Li gas is polarized into a single spin state, it is non-interacting and must be sympathetically cooled by Cs \footnote{Sympathetic cooling is a technique in which one atomic species is cooled indirectly through thermal contact with another species that is actively cooled, typically by forced evaporation \cite{myatt1997}. This method is essential for cooling atoms that cannot undergo evaporative cooling on their own, such as single-component Fermi gases with suppressed self-interactions.}. This is a challenge because the large mass ratio and differing scalar polarizability of Li and Cs both contribute to Cs sagging in the trap under gravity, potentially overlapping poorly with Li. To counteract the gravitational sag, we perform some of the final evaporation in a bichromatic trap consisting of both a 1064~nm laser (attractive for both species) and a 780~nm laser (attractive for Li, repulsive for Cs).

At the end of the preparation we have a Cs BEC immersed in a much larger degenerate Fermi gas of Li in an elongated geometry, see Fig.~\ref{fig:csli_image}.

\begin{figure*}[htbp]
\centering  
    \includegraphics[clip, trim=5cm 0cm 0.5cm 0cm, width=0.9\textwidth]{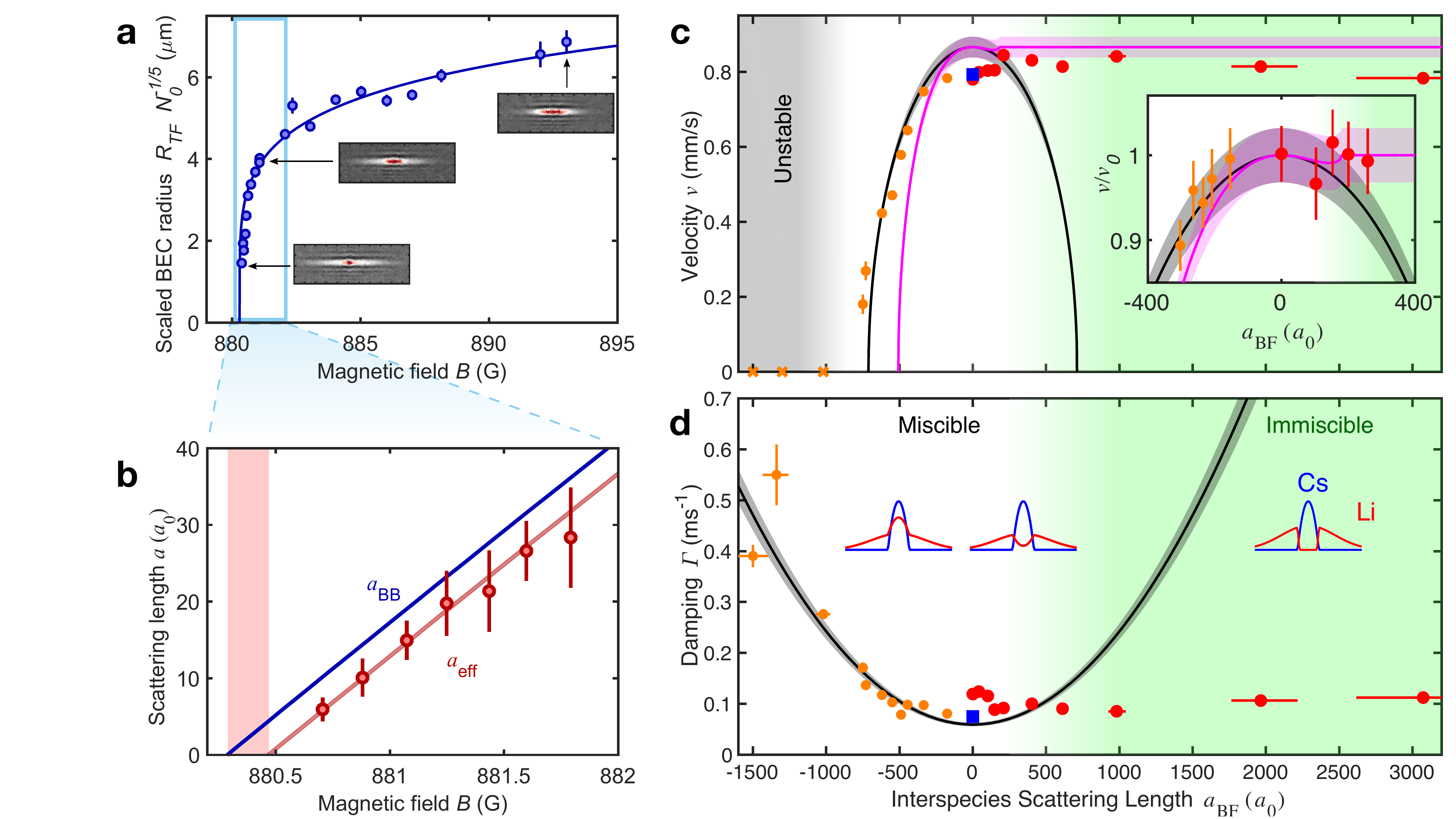}
    \caption{Experimental evidence of fermion-mediated interactions in a Cs BEC in the weak coupling regime. (\textbf{a}) \textit{In situ} Thomas–Fermi radius $R_\text{TF}$ of a Cs BEC as the Cs-Cs scattering length $a_\text{BB}$ approaches zero. Blue circles are measurements without Li, showing agreement with theoretical scaling (blue line). Insets: characteristic absorption images. (\textbf{b}) Relative measurements of $R_\text{TF}$ with and without Li extract the effective Cs-Cs scattering length $a_\text{eff}$ (red circles). A systematic negative shift indicates attractive mediated interactions. The red curve is a fit, and the blue line indicates the bare Cs-Cs scattering length $a_\text{BB}$. 
    Sound speeds and damping rates in the weak-coupling regime. (\textbf{c}) Density wave velocities at various interspecies scattering lengths $a_{\mathrm{BF}}$. Orange: attractive side ($a_{\mathrm{BF}}<0$); red: repulsive side ($a_{\mathrm{BF}}>0$); blue square: Cs-only reference. Crosses: samples without stable propagation. Inset: ratio of wave velocities with and without Li. Black/magenta lines: perturbative/mean-field theory \cite{yip2001,viverit2002,patel2023}. (\textbf{d}) Damping rates of the sound waves. Insets show schematic profiles of Cs and Li density in different regions of the scattering length $a_\text{BF}$. Shaded areas and error bars represent theoretical and experimental uncertainties in all plots. \textit{Source: (a, b) are adapted from Ref.~\cite{desalvo2019}, (c, d) are adapted from Ref.~\cite{patel2023}.}}  
    \label{fig:weak_interaction_combined}
\end{figure*}

\subsection{Weak-Coupling Regime}
\label{sec:lics:weak}

In the weak-coupling regime, where the interspecies scattering length $a_{\mathrm{BF}}$ is small compared to the interparticle spacing, the Cs BEC and Li Fermi gas coexist with minimal loss and strong spatial overlap. This regime provides a clean platform for exploring subtle interaction effects, such as fermion-mediated forces and induced mean-field shifts in the bosonic component. To probe these effects, we measure how the presence of the Li Fermi sea alters the static and dynamic properties of the Cs condensate. 

At weak interspecies attraction and small Cs-Cs repulsion, we observe changes in the spatial profile of the Cs BEC, in particular a reduction in its \textit{in situ} Thomas–Fermi radius $R_\text{TF}$ (Fig.~\ref{fig:weak_interaction_combined}a). These changes are attributed to attractive interactions induced by the surrounding fermions and are quantified in terms of an effective Cs–Cs scattering length $a_{\mathrm{eff}}$ in the presence of Li (Fig.~\ref{fig:weak_interaction_combined}b). The negative offset of $a_{\mathrm{eff}}$ relative to the bare scattering length $a_{\mathrm{BB}}$ is consistent with the perturbative, Born-approximation result given in Eq.~\eqref{eq:aind} \cite{desalvo2019,santamore2008}:
\begin{equation}\label{eq:aeff}
    a_\text{eff} = a_\text{BB} - \frac{k_\text{F}}{2\pi} \frac{(m_\text{B}+m_\text{F})^2}{m_\text{B}m_\text{F}}a_\text{BF}^2.
\end{equation}

We further investigate dynamic signatures of mediated interactions by studying collective excitations and density wave propagation in the Cs BEC. In particular, we observe shifts in dipole oscillation frequencies \cite{desalvo2019} and sound wave velocities \cite{patel2023}, as well as enhanced damping when Li is present. These features signal a shift of the bosonic excitation spectrum due to fermionic density fluctuations.

\begin{figure*}[htbp]
\centering  
    \includegraphics[clip, trim=25cm 0cm 0cm 0cm,width=0.5\textwidth]{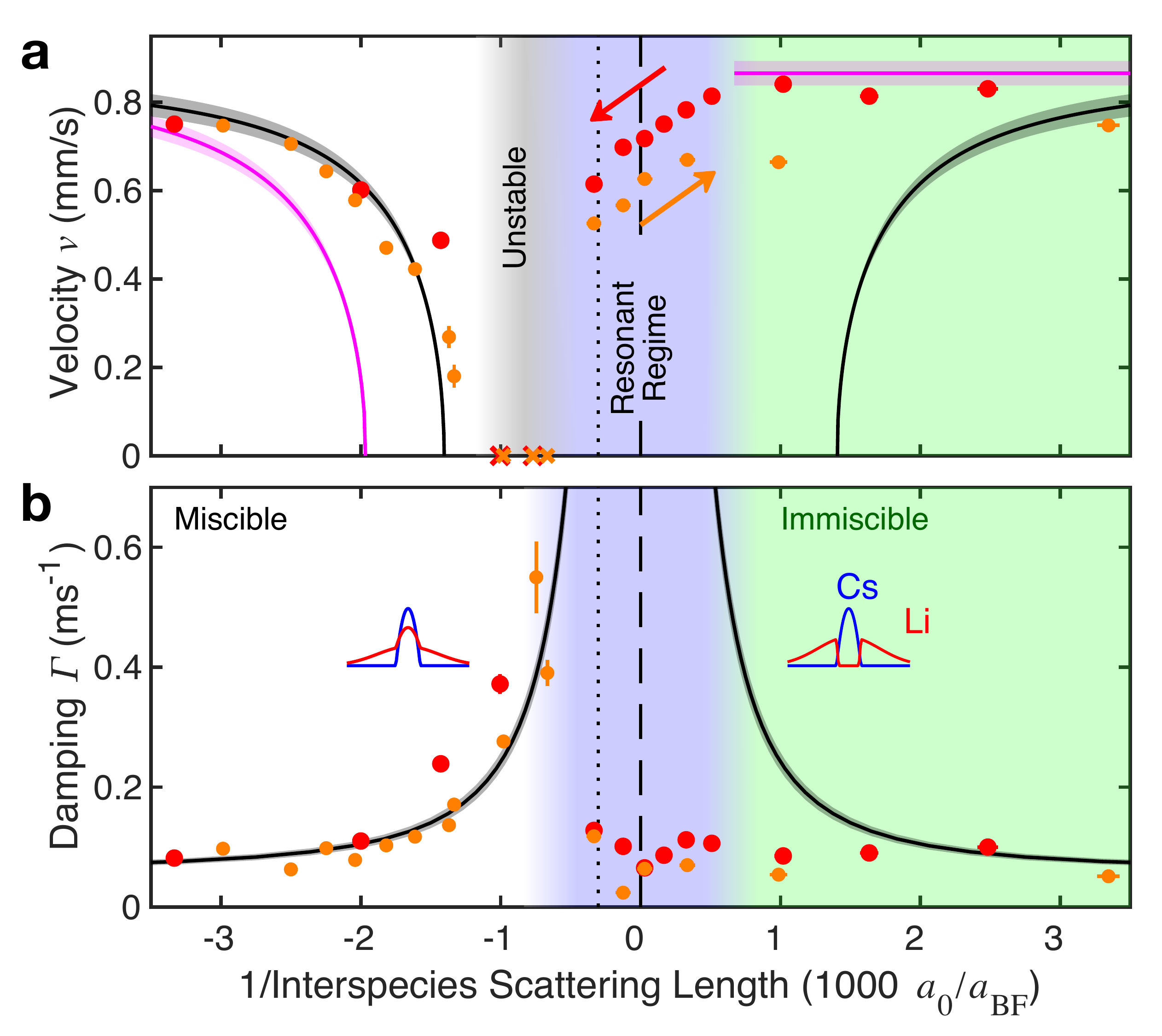}
    \caption{Sound propagation across the Feshbach resonance. (a) Density wave velocity as a function of $a_{\mathrm{BF}}$ shows nonmonotonic behavior, with the blue and green shaded regions marking the resonance and phase separation regimes. (b) Damping rates increase near resonance. The black and magenta curves are theoretical predictions from perturbative and mean-field models. Error bars and shaded regions denote experimental and theoretical uncertainties. Cartoons illustrate density profiles of Cs and Li with strong interspecies attraction and repulsion. The vertical dashed and dotted lines mark the Feshbach and Efimov resonance positions, respectively. \textit{Source: adapted from Ref.~\cite{patel2023}.}}
    \label{fig:strong_sound}
\end{figure*}

Figure~\ref{fig:weak_interaction_combined} summarizes our sound speed measurements. We disturb the Cs BEC locally with an optical potential and deduce the sound speed from the propagation of the density response. As the boson-fermion interaction becomes more attractive, the effective scattering length between Cs atoms decreases, see Eq.~\eqref{eq:aeff}, leading to a reduced sound wave velocity and an enhanced damping rate. In particular, the reduction in sound speed for weak boson-fermion attraction is consistent with the perturbative prediction Eq.~\eqref{eq:aeff} (black line). The flatness of the sound speed for interspecies repulsion is attributed to phase separation and matched well by a modified prediction that includes mean-field density effects (magenta line). These results provide a quantitative confirmation of induced interactions in a dilute Bose-Fermi mixture and illustrate the high sensitivity of the Cs BEC as a probe of the fermionic environment \cite{patel2023}.

Together, these observations demonstrate that our system can operate deeply in the coherent weak-coupling regime, where interactions between bosons can be engineered via fermion-mediated attractions. The ability to resolve small frequency shifts and extract effective interaction parameters underscores the experimental precision and stability of the platform.

\subsection{Strong-Coupling Regime}
\label{sec:lics:strong}

As the interspecies interaction strength increases near the Feshbach resonance, the system enters the strong-coupling regime, where the interspecies scattering length $a_{\mathrm{BF}}$ becomes comparable to or exceeds the interparticle spacing. In this regime, both few-body and many-body effects emerge prominently, and the system is no longer amenable to simple perturbative treatments.

A key question is how mediated interactions manifest in the strong coupling regime. In particular, can fermions mediate strong enough attraction to pair heavy bosons/impurities? While the RKKY potential (Eq.~\ref{eq:RKKY}) can support many bound states for large enough $a_{\mathrm{BF}}$, other theoretical models, summarized in Fig.~\ref{fig:aindth}, suggest only 0 or 1 resonance in the induced scattering length.

Experimentally, we observe significant modifications to the properties of the Cs BEC near the interspecies Feshbach resonance. First of all, the Cs superfluid sound speed becomes strongly dependent on the scattering length $a_{\mathrm{BF}}$, to the point where sound propagation fully stops for sufficiently strong interspecies attraction, indicating that the effective interaction between cesium atoms becomes negative (see Fig.~\ref{fig:strong_sound}a). This observation reflects a substantial change in the compressibility of the condensate, mediated by the dense surrounding Fermi sea \cite{patel2023}. Near resonance, sound propagation reappears, with a revival of collective modes in a narrow window. Concurrently, damping of the excitations increases, indicating enhanced collisional and decay processes (Fig.~\ref{fig:strong_sound}) \cite{patel2023}. The deviation from mean-field predictions in this regime points to emergent correlations and possible quasi-bound state formation.

\begin{figure}
\centering  
    \includegraphics[clip, trim=39cm 0cm 0cm 0cm,width=0.9\linewidth]{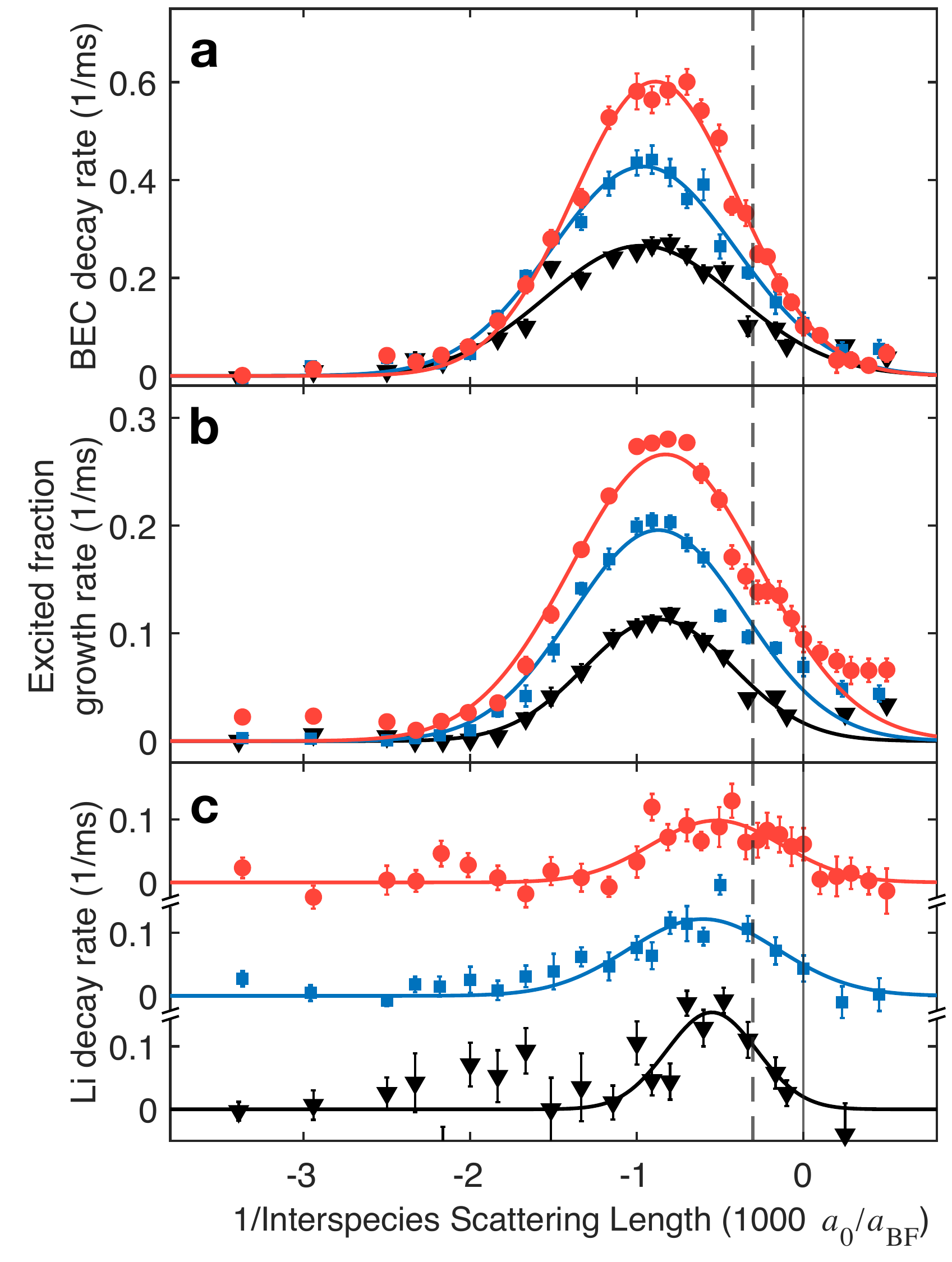}
    \caption{Resonant decay and excitation of Cs BECs induced by the degenerate Li Fermi gas. (\textbf{a-c}) Resonances in Cs BEC decay rate, Cs excitation fraction, and Li atom loss are observed at consistent values of $a_{\mathrm{BF}}$, distinct from known Efimov features. Red circles, blue squares, and black triangles indicate experiments with 20,000, 10,000, and 5,000 Li atoms, respectively. BEC and excitation fractions are determined by bimodal fits to \textit{in situ} images. Decay rates come from exponential fits. Error bars show standard deviations of the mean. Vertical lines indicate the Feshbach (solid) and Efimov (dashed) resonance positions. \textit{Source: adapted from Ref.~\cite{cai2025}.}}
    \label{fig:mediated_resonance}
\end{figure}

A particularly striking observation is the appearance of a previously unidentified resonance structure in the Cs BEC loss and heating spectra. This resonance occurs on the attractive side of the interspecies Feshbach resonance and is distinct from the known Efimov three-body resonance at $a_\text{BF} = -3300\,a_0$ \cite{johansen2017}. As illustrated in Fig.~\ref{fig:mediated_resonance}, this feature manifests as a sharp enhancement in Cs BEC depletion, excitation growth, and Li atom loss. The position of the resonance depends weakly on the Li atom number, suggesting a collective or many-body origin \cite{cai2025}. The resonance is interpreted as a fermion-mediated boson-boson pairing resonance — a type of effective interaction not present in purely bosonic or fermionic systems. Its appearance signals a new mechanism of correlation-driven binding enabled by the mass and density asymmetry of the Cs–Li mixture.

These observations provide compelling evidence for the emergence of novel quantum states in the strongly interacting regime. The interplay between induced interactions, three-body loss, and pairing phenomena opens new pathways for exploring unconventional superfluidity, composite particles, and quantum critical behavior in ultracold mixtures.

\newpage

\section{Outlook}
\label{sec:out}

In this chapter, we have outlined the theoretical framework and results for mediated interactions in ultracold atomic mixtures, along with experimental studies of fermion-mediated interactions in $^6\text{Li}$–$^{133}\text{Cs}$ mixtures. In the weak-coupling regime, our measurements show good agreement with theoretical predictions. Building on these advances, several promising but unexplored directions remain open for experimental investigation.

An important next step is to verify the long-range nature of the mediated interactions, which will greatly expand the scope of quantum simulation. The predicted interaction range $1/k_\text{F} \approx 1~\mu$m, is comparable to a typical optical lattice spacing. Loading the bosonic atoms into an optical lattice could make such effects observable as non-local, nearest-neighbor interactions between localized bosons \cite{de2014,arguello2022}.

Another intriguing possibility arises when the fermion-mediated attraction exceeds the boson–boson repulsion, potentially leading to self-bound Bose–Fermi droplets stabilized by beyond-mean-field effects \cite{rakshit2019}. A key challenge is to understand the role of three-body processes in both mechanical stability and collisional loss. Such Bose-Fermi droplets bring analogies to astronomical objects like white dwarfs or neutron stars \cite{rakshit2019}, and so their realization in cold atom systems would be of great interest to quantum simulation.

A third avenue involves exploring mediated interactions within BCS–BEC mixtures—systems containing a BEC coexisting with a fermionic superfluid in the BCS–BEC crossover \cite{zheng2024mediated}. Such mixtures have been realized in several atomic species combinations \cite{ferrier2014,ikemachi2016,yao2016,roy2017}, and the $^{133}\text{Cs}$–$^6\text{Li}$ system is a particularly promising candidate thanks to its convenient Feshbach tuning. As illustrated in Fig.~\ref{fig:feshbach}, there exists a magnetic-field window where a stable BEC of Cs coexists with strongly attractive Li fermions, while Li–Cs resonances allow tunable interspecies coupling.

Finally, we note that pioneering work on mediated interactions in other ultracold mixtures has demonstrated a wide range of rich phenomena, including spin-spin fermion mediated interactions \cite{edri2020}, suppression of unitary three-body loss due to fermion mediated interactions \cite{chen2022}, and rich polaron physics \cite{fritsche2021,baroni2024mediated}. Taken together, theoretical advances and experimental capabilities have positioned the field to enter a new era, where the controlled engineering of mediated interactions can be harnessed to explore exotic quantum matter, realize novel many-body phases, and deepen our understanding of strongly correlated systems. The prospects ahead are both diverse and within experimental reach.

\section{Acknowledgements}
This chapter will appear in the Springer book entitled ``Short and Long Range Quantum Atomic Platforms - Theoretical and Experimental Developments.''
T.E. thanks Olivier Bleu for inspiring discussions. We thank Brian DeSalvo, Krutik Patel, and Sarah McCusker for their contributions to the experiments. The Heidelberg work was supported by the DFG (German Research Foundation) under Project No.\ 273811115 (SFB 1225 ISOQUANT) and under Germany’s Excellence Strategy EXC2181/1-390900948 (the Heidelberg STRUCTURES Excellence Cluster). The University of Chicago work was supported by the National Science Foundation under Grant No.\ PHY-2409612 and by the Air Force Office of Scientific Research under Award No. FA955021-1-0447. H.A. acknowledges support by the National Science Foundation Graduate Research Fellowship under Grant No.\ DGE 1746045.

% BibTeX users please use
\bibliographystyle{apsrev_titles}
\bibliography{all}

\end{document}